# Electrical transport properties of bulk $MgB_2$ materials synthesized by the electrolysis on fused mixtures of $MgCl_2$, NaCl, KCl and $MgB_2O_4$

Kenji Yoshii and Hideki Abe*

*Japan Atomic Energy Research Institute, Mikazuki, Hyogo 679-5148, Japan*
*\*National Institute for Materials Science, Tsukuba, Ibaraki 305-0047, Japan*

Kenji Yoshii
Japan Atomic Energy Research Institute, Mikazuki, Hyogo 679-5148, Japan,
Fax: 081-791-58-2740
e-mail: yoshiike@spring8.or.jp

Electrolysis was carried out on fused mixtures of $MgCl_2$, NaCl, KCl and $MgB_2O_4$ under an Ar flow at 600°C. Electrical resistivity measurements for the grown deposits show an onset of superconducting transition at ~37 K in the absence of applied magnetic field. The resistivity decreases down to zero below ~32 K. From an applied-field dependence of resistivity, an upper critical field and a coherence length were calculated to be 9.7 T and 5.9 nm at 0 K, respectively.

Recent discovery of superconductivity in $MgB_2$ at a transition temperature ($T_C$) as high as 39 K [1] has shown a possibility of new superconducting materials for low-cost and high-performance electrical devices and magnets [2-14]. This transition temperature much exceeds the highest value of binary intermetallic compounds, $T_C$=23 K for the A15-type compound $Nb_3Ge$ [15]. Only the high- $T_C$ cuprates and alkaline-doped $C_{60}$ are known to have the comparable order of $T_C$ values [16]. From the viewpoint of industrial application, wires [3,11], tapes [12] and thin films [13] as well as bulk samples [9,10,14] were fabricated by means of direct reactions between elemental Mg and B.

The present authors have recently proposed an electrochemical synthesis method of $MgB_2$ [17]. Black deposits containing $MgB_2$ were obtained by means of electrolysis on a fused mixture of $MgCl_2$, KCl and $MgB_2O_4$ with a molar ratio of 5:5:1 (denoted as KCl-bath for brevity hereafter) at the temperatures lower than the melting point of Mg, 651°C. The simple installation, moderate reaction condition and the low-cost starting materials used in this preparation method are favorable for practical application. Though a superconducting transition was confirmed by magnetic measurements, a resistive superconducting transition has not been observed due to high contact resistance of the samples. This is because the deposits prepared include considerable amount of the insulating electrolyte [17].

In order to improve the electrical transport properties of the deposits, effects of substitution of KCl by the other alkaline halides, such as LiF, LiCl, NaF and NaCl, were investigated in the present paper. It was found that electrical resistivity measurements could be carried out for the deposits obtained from a NaCl-based electrolyte. The upper critical field ($H_{C2}(T=0)$) and the coherence length ($\xi_0$) were evaluated from the field dependence of electrical resistivity.

The installation of the electrolysis was the same as that reported in the previous work [17]. An alumina boat equipped with a graphite anode and a Pt-wire cathode wasÅ@used as the electrolysis cell, in which electrolysis for the molten electrolyte was carried out at an applied voltage of 5 V at 600°C. Contents ($x$) of the alkaline halides ($AX$) were changed between 0 and 10 in the molar ratio of $MgCl_2$, $AX$, KCl, and $MgB_2O_4$, 10: $x$: (10-$x$): 2. As the lowest contact resistance has been achieved for $x$=7 with $AX$=NaCl so far, experimental results will be discussed only for the samples prepared from this electrolyte (abbreviated as the NaCl-bath). The melting points of the NaCl- and KCl-baths were 450°C and 580°C, respectively, which were estimated from ionic current measurements during the electrolysis. The currents were an order of a few tens milli-ampere for both baths.

Figure 1 shows the close-up views of the electrolysis cells at around the Pt anodes after the electrolysis of 30 minutes for the KCl-bath (Fig.1(a)) and of the NaCl-bath (Fig.1(b)). The boundaries between the deposits (dark area) and the insulating electrolyte (bright area) in Fig.1(a) are clearer than that in Fig.1(b). This difference suggests that the deposits prepared from the NaCl-bath contain fewer amounts of the insulating electrolyte than the deposits from the KCl-bath. Contact resistance of the former deposits was found to be lower by decades than that of the latter deposits, allowing the electrical resistivity measurements with considerable reliability.

Figure 2 shows electrical resistivity plotted as a function of temperature for the deposits of the NaCl-bath, which was measured with different applied magnetic fields. It was found that the resistivity decreases monotonically with decreasing temperature. A residual resistivity ratio (RRR) defined as $\rho$(300 K)/$\rho$(40 K) ($\rho$: resistivity) was calculated to be 2.23 at the zero magnetic field. The normal state resistivity around 40 K is higher than that of bulk $MgB_2$ reported in previous papers [3-7,9,10], which implies a weak linkage of conducting domains.

An onset of superconducting transition is observed at ~37 K in the absence of applied field. This temperature is close to that reported in Ref. [17]. The resistivity drops down to less than $10^{-4}$ ohm cm below ~32 K. Figure 2 also shows monotonic broadening of the resistive superconducting transition with increasing applied magnetic field as was reported in several papers [5,6,9,10,18-20], which is interpreted in connection with the weak linkage of superconducting domains [5]. The upper critical field $H_{C2}(T=0)$ ($T$: temperature) was estimated from the applied-field dependence of resistivity. The transition fields defined at the half values of the normal-state resistivity were plotted as a function of temperature in Fig.3. Simple linear extrapolation shown in this figure provides $H_{C2}(T=0)$ of ~12 T. Assuming the relationship of $H_{C2}(T=0)=0.73*T_C*[-dH_{C2}(T)/dT]$ for an $s$-wave superconductor [20], a value of $H_{C2}(T=0)$ is calculated to be 9.7 T. The previously reported $H_{C2}(T=0)$ values range from 13 to 18 T for the polycrystals [18,19] and from 5 to 20 T for the single crystals [5,6,20]. It is qualitatively relevant that the present value of 9.7 T stands between those determined for the single-crystal in the *ab*- plane (15 T) and along the *c*-axis (4.8 T) [20]. The $H_{C2}(T=0)$ of 9.7 T leads to a coherence length at 0 K ($\xi_0$) of 5.9 nm by use of the Ginzburg-Landau equation $\xi_0 = [\phi_0/2\pi H_{C2}(T=0)]^{1/2}$ ($\phi_0$: flux quantum) [18]. The coherence lengths are reported to be $\xi_0 \sim$ 6-8 nm in the *ab*-plane and $\xi_0 \sim$ 2-3 nm along the *c*-axis for the single-crystals [6,20]. The present $\xi_0$ value of 5.9 nm is comparable to that obtained for a polycrystal, 4.7 nm [18].

In summary, the samples obtained from the electrolysis on fused mixtures of $MgCl_2$, NaCl, KCl and $MgB_2O_4$ were found to exhibit the resistive superconducting transition at the onset temperature of ~37 K and the zero-resisitivity below ~32 K. From the applied-field dependence of resistivity, the upper critical field and the coherence length were calculated to be 9.7 T and 5.9 nm at 0 K, respectively. The observation of the resistive superconducting transition indicates that the present synthesis method offers a route to fabrication of realistic superconducting magnets and devices. Further investigations are currently in progress, and their results will be published in the near future.

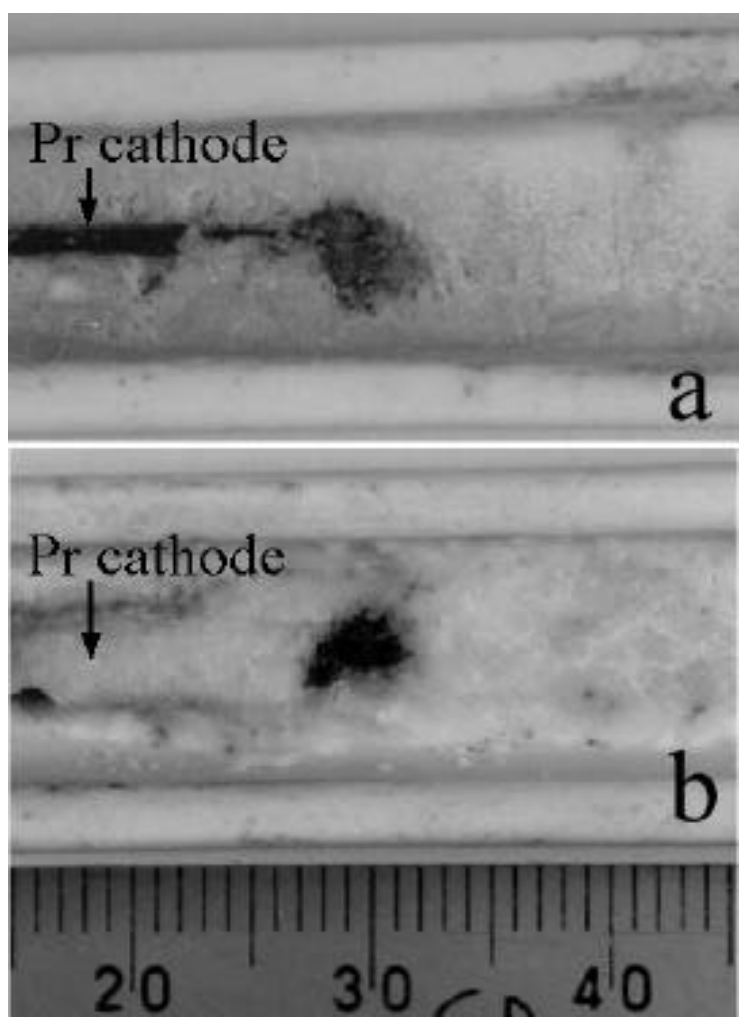

FIG.1. Photograph of deposits (dark areas around the centers) prepared from (a) KCl-bath and (b) NaCl-bath. The scale in unit of mm is shown at the bottom.

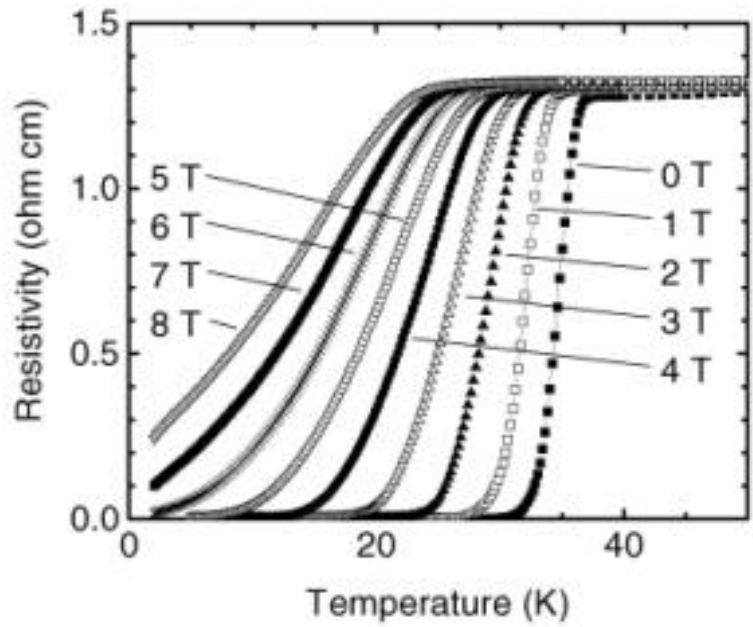

FIG.2. Electrical resistivity plotted against temperature for the sample prepared from the NaCl-bath. Applied magnetic fields were changed between 0 and 8 T with the increment of 1 T, each of which is shown in the figure.

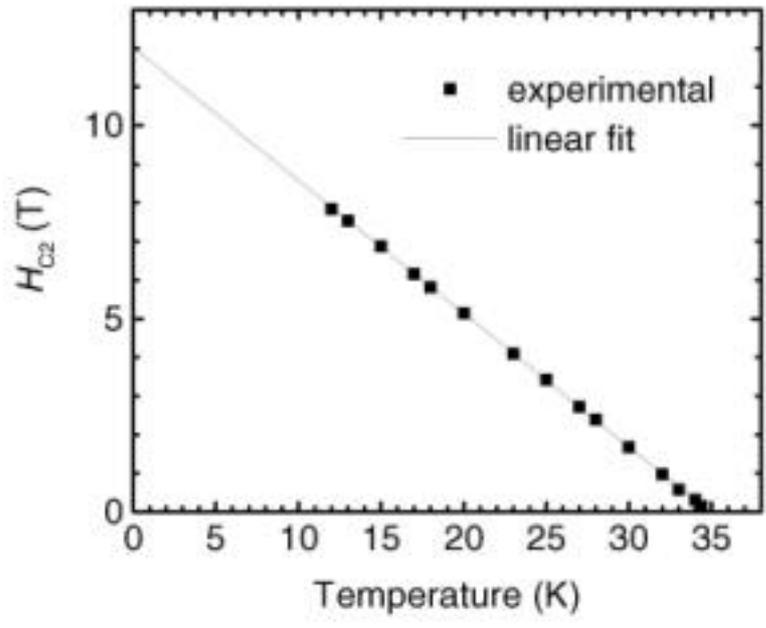

FIG.3. Upper critical field ($H_{C2}$) plotted against temperature ($T$), which was determined from the resistivity data (FIG. 2). Linear extrapolation of the experimental data is shown as linear fit.